\documentclass[8 pt]{article}

\usepackage{amssymb}
\usepackage{amsfonts}
\usepackage{amsmath}
\usepackage{perpage}
\usepackage{epigraph}
\MakePerPage{footnote}
\usepackage[titletoc,title]{appendix}

\begin{document}

\title{A Covariant Approach to Entropic Dynamics\footnote{Presented at MaxEnt 2016, the 36th International Workshop on Bayesian Inference and
Maximum Entropy Methods in Science and Engineering (July 10–15, 2016, Ghent, Belgium).}}

\author{Selman Ipek, Mohammad Abedi, Ariel Caticha\footnote{sipek@albany.edu, mabedi@albany.edu, acaticha@albany.edu}\\
{\small Physics department, University at Albany - SUNY, Albany, NY, 12222, USA}}
\date{}

\maketitle

\begin{abstract}
Entropic Dynamics (ED) is a framework for constructing dynamical theories of inference using the tools of inductive reasoning. A central feature of the ED framework is the special focus placed on time. In \cite{Caticha 2013}\cite{Caticha Ipek 2014} a global entropic time was used to derive a quantum theory of relativistic scalar fields. This theory, however, suffered from a lack of explicit or manifest Lorentz symmetry. In this paper we explore an alternative formulation in which the relativistic aspects of the theory are manifest.

The approach we pursue here is inspired by the works of Dirac, Kucha\v{r}, and Teitelboim in their development of covariant Hamiltonian methods. The key ingredient here is the adoption of a local notion of time, which we call entropic time. This construction allows the expression of arbitrary notion of simultaneity, in accord with relativity. In order to ensure, however, that this local time dynamics is compatible with the background spacetime we must impose a set of Poisson bracket constraints; these constraints themselves result from requiring the dynamcics to be \textit{path independent}, in the sense of Teitelboim and Kucha\v{r}.
\end{abstract}

\section{Introduction}
ED is a framework for constructing dynamical theories of physics from the principles of inference. The goal is not modest: the ED program aspires to demonstrate that the laws of physics are a natural consequence of the rules for processing information, i.e. probability theory and the method of maximum entropy (ME). The primary task is to update the probability $\rho$, subject to the appropriate information, that is, subject to the appropriate constraints --- this is where the appropriate physics is incorporated into the theory.

There are several manners in which one constrains the theory to reflect the relevant physics. One such constraint is that the probability $\rho$ be guided by a drift potential $\phi$. In turn, one must supply an additional constraint that dictates the update of $\phi$ in response to the changing $\rho$. In short, ED is dynamical theory that is driven by constraints.

Another place where constraints arise, is how time appears in the theory. Indeed, while the rules of inference are themselves atemporal --- we can be uncertain about the past, present, or future --- dynamics itself is closely intertwined with time. Thus in ED, a dynamical theory of inference, time must be constructed by defining a notion of an instant, how such instants are ordered, and the duration between them. However, any such construction must also be sure to respect the physics we wish to express, hence constraining the theory.

This constraint-centric approach to physics and time, as advocated by ED, has been successful in deriving many aspects of quantum theory. For example, in \cite{Bartolomeo et al 2014}, a quantum theory of non-relativistic particles in flat space was developed. The derivation hinged on two crucial points. The first was the implementation of a global notion of an instant, in the Newtonian style, in which all the degrees of freedom at all spatial points were updated equally; a constraint that is reasonable in the context of a flat space. Moreover, once one adopts a global notion of time, such as this, it becomes natural to introduce the notion of a \textit{global} energy function; we then impose that the update of $\phi$ be determined by requiring this energy to \textit{conserved}.\footnote{The notion that a conserved energy is the key constraint to implement for a non-dissipative quantum dynamics was first recognized Nelson in \cite{Nelson}, and was first introduced in ED with \cite{Caticha 2010a}.} The result is a quantum ED that also has the structure of a Hamiltonian dynamics.

What is perhaps surprising, is that a similar model of ED was utilized in \cite{Caticha 2013}\cite{Caticha Ipek 2014} for the purposes of deriving a theory of \textit{relativistic} quantum scalar fields in the Sch\"{o}dinger picture. That is, the notion of a global Newtonian instant and energy conservation --- which was sufficient for deriving a non-relativistic quantum theory --- also turned out to be successful in the relativistic domain. This situation is reminiscent of the status of quantum field theory in the early 1930's, shortly after the advent of the new quantum mechanics by Heisenberg, Schr\"{o}dinger, and Dirac.\footnote{The biographical book by Silvan Schweber \cite{Schweber} provides, in chapter one, an excellent overview of the status of quantum field theory at that time.} At that time, physicists were borrowing wholesale the quantization methods developed for non-relativistic particles, for the purposes of relativistic quantum theories. The problem encountered in these early relativistic approaches, and which is similar in spirit to the problems encountered in \cite{Caticha 2013}\cite{Caticha Ipek 2014}, is that, although these theories were relativistic in content, they were not explicitly so. That is, the freedom to represent the different notions of simultaneity, typically associated with relativity, is absent --- the theory is not \textit{manifestly} covariant.\footnote{One glaring example of this was the non-covariant procedure of quantizing via the ``equal-time commutation relations," which specified a preferred notion of time.}

Work done to overcome this problem in the standard quantum theory began with the early \textit{many-time} efforts of P. Weiss \cite{Weiss 1938} and Dirac \cite{Dirac 1932}. These ideas were later further developed by Dirac \cite{Dirac 1948}, and were adopted by others, such as S. Tomonaga \cite{Tomonaga 1946} and J. Schwinger \cite{Schwinger 1948 I} in their approaches to a fully covariant quantum theory. (Their ability to do so contributed to their Nobel prize (1965) winning insights into quantum electrodynamics.) Collectively, these initial works could be characterized as precursors for the subsequent covariant Hamiltonian methods developed, again, by Dirac \cite{Dirac Lectures}\cite{Dirac 1951}, but which were later refined by K. Kucha\v{r} \cite{Kuchar 1972} and C. Teitelboim \cite{Teitelboim thesis}\cite{Teitelboim 1972} in their pursuit of a covariant approach to geometrodynamics.

Here we draw specifically on the covariant methods of Dirac, Kucha\v{r}, and Teitelboim (DKT) for the purposes of developing a relativistically covariant formulation of ED. Following DKT, we slice, or foliate, the background spacetime by a sequence of spatial cuts --- called hypersurfaces, or more simply, just \textit{surfaces}. While this process obscures the original local four-dimensional Lorentz symmetry of spacetime, it can be recovered by the implementation of a new symmetry: foliation invariance. This foliation symmetry is then, in turn, implemented by the kinematic requirement of embeddability \textit{\`{a} la } Teitelboim and Kucha\v{r}: if a final surface can be reached from an initial surface by a number of intermediate paths, then all such paths should agree.

To implement this scheme in ED we must first identify, in the context of a relativistic ED, what is meant by an instant of time. An instant in ED is defined as consisting of two main ingredients; one kinematic, the other, epistemic. The first ingredient is a spacelike surface in spacetime; this allows us to express our desired notion of simultaneity and is this key step that allows for the notion of a relativistic instant, in deviation from \cite{Caticha 2013}\cite{Caticha Ipek 2014}. And to this surface we attach, in addition, our second ingredient, the epistemic state $\rho$ and $\phi$ --- this is the information that we wish to monitor and update.

Having established the notion of an instant, we can envision dynamics as an update of the state $\rho$ and $\phi$ from instant to instant. However, just as one curved surface is deformed into another in a local fashion, the evolution of $\rho$ and $\phi$ must follow suit. That is, we must also implement a local updating scheme for evolving $\rho$ and $\phi$. This is achieved by the introduction of a set of local time parameters, one per spatial point, for the purposes of what we call a local time entropic update. And while the probability $\rho$ is, naturally, still updated using the standard entropic methods, we can no longer update $\phi$ by requiring a conserved energy, as such a concept is typically not available in a general spacetime. Instead, drawing on the work of Teitelboim and Kucha\v{r}, we propose here that the update of $\phi$ be determined by requiring the joint dynamics of $\rho$ and $\phi$ to be \textit{path independent}: if there are many ways to evolve between a given initial and final state, then all ways should agree.

Executing this idea, however, requires a choice of language, or \textit{representation}. That is to say, the requirement of path independence is expressed as a set of local brackets, three per spatial point, which require a specific framework for implementing them. Here we express the requirement of path independence by adopting a minimal Hamiltonian formalism, where $\rho$ and $\phi$ play the role of conjugate variables. The requirement of path independence can then be phrased as a collection of Poisson bracket relations to be satisfied by a set of Hamiltonian generators. The resulting covariant ED leads naturally to a set of local time Schr\"{o}dinger equations, and thus we claim, to a covariant quantum theory.

This covariant ED scheme is implemented here for the case of a single scalar field in an arbitrary background spacetime. In section \ref{Entropic Dynamics} we outline our model for inferring the dynamics of the scalar field. In section \ref{Entropic time} we construct a notion of time to assist in organizing our inferences. Section \ref{Spacetime and consistency} introduces the kinematic notion of embeddability and the dynamical principle of path independence, while in section \ref{Non-dissipative covariant dynamics} we utilize these concepts for updating $\phi$, which results in a quantum theory. In section \ref{Conclusion} we discuss our results.
\section{Entropic Dynamics}
\label{Entropic Dynamics}
Our goal here is to predict and understand the dynamical behavior of a scalar field. We proceed by making two main dynamical assumptions. The first is that change happens; we do not ask why, we only set out to proceed rationally, given that it, in fact, does. Our second assumption is that, in ED, motion is continuous, such that large changes follow from the accumulation of many infinitesimally small steps. The ED framework provides a methodology for estimating this infinitesimal dynamics.
\paragraph{Microstates and notation}
In an inference scheme the physics is introduced through
the choice of variables and of constraints. Here we deal with a scalar field $\chi \left( x^{i}\right)\equiv \chi _{x}$, which has values that are definite, but uncertain; it is these values that we wish to infer. The field $\chi_{x} $ lives on a three dimensional surface $\sigma$, which is labeled by the coordinates $x^{i}$ ($i=1,2,3$) and is, by definition, a scalar with respect to three-dimensional coordinate transformations of the surface. Moreover, this surface $\sigma$ is endowed with a metric $g_{ij}(x)$, projected from the surrounding spacetime. This allows us to properly define integration measures using the determinant of the metric, $g_{x}^{1/2}=\det\left\vert g_{ij}\right\vert ^{1/2}$.

The goal is to determine the likelihood of a particular field configuration $\chi_{x}$. That is, we wish to determine the probability $P(\chi_{x_{1}} = \chi_{1}, \, \chi_{x_{2}}=\chi_{2},\cdots)$, for each point $x\in \mathbb{R}^{3}$, that the field variable $\chi_{x_{1}}$ labeled by the position $x_{1}$ has value $\chi_{1}$, and so on. A more convenient description is afforded by the use of a configuration space $\mathcal{C}$, which is the infinite dimensional space of all possible field configurations. A single field configuration, labeled $\chi = \{\chi_{x_{1}}, \chi_{x_{2}},\cdots\} $, is represented as a point in $\mathcal{C}$. Thus, the uncertainty in the field can be conveniently encoded through probability distributions, such as $P[\chi]$, over $\mathcal{C}$.\footnote{A comment on notations. We use here a notation that largely argrees with \cite{Caticha 2013}\cite{Caticha Ipek 2014}. Functional dependence $f:\mathcal{C}\to\mathbb{R}$ is denoted by square brackets, such as $f[\chi]$. As is common, functional derivatives with respect to a particular field degree of freedom $\chi_{x}$ is given by $\delta/\delta\chi_{x}$. We employ the practice of denoting functional integration measures $D\chi\sim\prod_{x}d\chi_{x}$. Additional notation is introduced as needed.}

\paragraph{Maximum Entropy}
We are dealing with a scalar field $\chi(x)$ where we are uncertain of the values of $\chi$ and what subsequent values $\chi^{\prime}$ we expect it will attain. The appropriate probability distribution that quantifies this informational state is the joint distribution $P[\chi^{\prime},\chi] = P[\chi^{\prime}|\chi]\,\rho[\chi]$; where we have used the product rule of probability theory. The typical dynamical problem we wish to solve is that, given some initial information about our system, what future behavior do we expect? The distribution that addresses this issue is the transition probability $P[\chi^{\prime}|\chi]$, which gives the probability that the field makes a small step from a configuration $\chi\in \mathcal{C}$ to an unknown $\chi^{\prime}\in \mathcal{C}$; our main interest here is thus determining a suitable $P[\chi^{\prime}|\chi]$.

At the outset of our problem, before we have learned anything, our knowledge of the dynamics is quantified by the so-called \textit{prior} distribution $Q[\chi^{\prime}|\chi]$. The distribution $Q[\chi^{\prime}|\chi]$ that reflects our maximally ignorant state of affairs is one that neglects correlations between the degrees of freedom, and which is unconstrained in the possible $\chi^{\prime}$'s. An appropriate choice is a uniform prior distribution,\footnote{How one handles prior information is in general a contentious topic. While here we have chosen $Q[\chi^{\prime}|\chi]$ to be uniform (or, more rigorously, a sufficiently broad Gaussian), there are other manners for proceeding, none of which affect our results. In short: ED is rather robust with respect to the choice of prior.}
\begin{equation}
Q[\chi^{\prime}|\chi] = \prod_{x} Q(\chi_{x}^{\prime}|\chi_{x})\sim Q.
\end{equation}

That such a prior distribution is insufficient for making predictions is obvious. To proceed, we must obtain more information. In ED, just as in any inference problem, information is supplied in the form of constraints. The method for choosing a suitable posterior distribution $P[\chi^{\prime}|\chi]$ that satisfies these constraints is provided by the method of Maximum Entropy (ME) \cite{Caticha 2004}\cite{Caticha 2012}.

The non-trivial step in this process is the identification of the relevant information for updating. Here the information that is suitable to our task is, first, that motion be continuous. This is imposed by the constraint
\begin{equation}
\left\langle \left( \Delta \chi _{x}\right) ^{2}\right\rangle \equiv \int
D\chi ^{\prime }\,P\left[ \chi ^{\prime }|\chi
\right] \left( \Delta \chi _{x}\right) ^{2}=\Delta\kappa _{x},
\label{Constraint 1}
\end{equation}%
which is a quadratic constraint for every spatial point $x$. The $\Delta\kappa _{x}$ are required to be small quantities, and in the limit $\Delta\kappa _{x}\to \infty$, will ensure that the motion is continuous.

On its own, the constraints (\ref{Constraint 1}) lead to a bland dynamics where the field variables exhibit no correlations. A richer dynamics is achived by imposing, in addition, a single \textit{global} constraint
\begin{equation}
\left\langle \Delta \phi\right\rangle =\int D\chi ^{\prime }\,\int dx\,\,P\left[ \chi ^{\prime }|\chi
\right]  \,\Delta \chi _{x}%
\frac{\delta \phi \left[ \chi \right] }{\delta \chi _{x}}=\Delta\kappa ^{\prime }.
\label{Constraint 2}
\end{equation}
where the constant $\Delta\kappa ^{\prime }$ also tends toward zero. The quantity $\phi$ is called the drift potential and it plays the role of a pilot wave that guides our inferences, resulting, eventually, in quantum effects such as entanglement and interference.\footnote{Notice that while $\chi _{x}$ and thus $\Delta \chi _{x}$ are scalars by definition, for (\ref{Constraint 2})
to be properly defined under coordinate transformations of the surface, it must be that $%
\delta \phi /\delta \chi _{x}$ transforms as a scalar density.}

To determine the transition probability $P\left[ \chi ^{\prime }|\chi
\right]$ we maximize the entropy,
\begin{equation}
S\left[ P,Q \right] =-\int
D\chi ^{\prime }P\left[ \chi ^{\prime }|\chi\right]  \log \frac{P\left[ \chi ^{\prime }|\chi\right]  }{Q\left[ \chi ^{\prime }|\chi\right]  },
\label{Entropy Transition Probability}
\end{equation}%
subject to the constraints (\ref{Constraint 1}) and (\ref{Constraint 2}),
normalization, and with respect to the prior introduced above.

Maximizing the entropy (\ref{Entropy Transition Probability}) with respect
to the constraints leads to a Gaussian transition probability distribution%
\begin{equation}
P\left[ \chi ^{\prime }|\chi\right] =\frac{1}{Z%
\left[ \alpha _{x},g_{x}\right] }\,\exp -\frac{1}{2}\int dx\,g_{x}^{1/2}\alpha _{x}\left( \Delta \chi _{x}-\frac{1}{g_{x}^{1/2}\alpha _{x}}\frac{\delta \phi \left[ \chi \right] }{\delta \chi
_{x}}\right) ^{2},  \label{Trans Prob}
\end{equation}%
where $Z\left[ \alpha _{x},g_{x}\right] $ is a normalization constant, the Lagrange multiplier $\alpha _{x}\rightarrow \infty $ is a scalar quantity which enforces continuity of the motion. In previous work $\alpha_{x}$ was chosen to be a spatial constant $\alpha$ to reflected the translational symmetry of the theory. Here we make no such restriction and instead relax the assumption of a global $\alpha$ in favor of a non-uniform spatial scalar $\alpha_x$. This will be a key element in implementing our scheme for a local entropic time.\footnote{The Lagrange multiplier $\alpha^{\prime}$ associated with the global constraint (\ref{Constraint 2}) turns out to be unimportant and can be chosen so that $\alpha^{\prime}=1$. The interested reader can look to \cite{Bartolomeo/Caticha 2016}\cite{Caticha/Cararra 2017} for more information.}
\paragraph*{Motion}
The Gaussian form of (\ref{Trans Prob}) allows us to present a generic
change $\Delta \chi _{x}=\left\langle \Delta \chi _{x}\right\rangle +\Delta
w_{x}$ as resulting from an expected drift $\left\langle \Delta \chi
_{x}\right\rangle $ plus fluctuations $\Delta w_{x}$. The expected step is 
\begin{equation}
\left\langle \Delta \chi _{x}\right\rangle =\frac{1}{g_{x}^{1/2}\,\alpha _{x}}\frac{\delta \phi \left[ \chi \right] }{\delta \chi_{x}}\equiv \Delta \bar{\chi}_{x},  \label{Exp Step 1}
\end{equation}%
while the fluctuations $\Delta w_{x}$ satisfy,%
\begin{equation}
\left\langle \Delta w_{x}\right\rangle =\left\langle \Delta \chi
_{x}-\left\langle \Delta \chi_{x}\right\rangle \right\rangle =0,\hspace{.5 cm}\textrm{as well as,}\hspace{.5 cm}
\left\langle \Delta w_{x}\Delta w_{x^{\prime }}\right\rangle
=\frac{1}{g_{x}^{1/2}\alpha _{x}}\delta _{xx^{\prime }}.
\label{Fluctuations}
\end{equation}%
Thus we see that even as $\Delta \bar{\chi}_{x}\sim 1/\alpha _{x}$, the fluctuations $\Delta w_{x}\sim 1/\alpha _{x}^{1/2}$; for $\alpha
_{x}\rightarrow \infty $ the fluctuations dominate the motion, which leads to a continuous, but non-differentiable trajectory --- a Brownian motion.

\section{Entropic time}
\label{Entropic time}
Having determined what short steps we are likely to expect, our next task is to devise a scheme for iterating these shorts steps and record their progression. Doing this require that we introduce a notion of time. We call this scheme \textit{entropic time}.\footnote{A more in depth discussion of entropic time is given in \cite{Caticha 2010a}.}
\paragraph{what is an entropic instant?}
In ED time is constructed as a device for organizing and monitoring the status of our inferences. As with any tool, its purpose is pragmatic in nature; that is, it is built to accomplish a particular task. Here we introduce a notion of entropic time that is built to express time in a fashion that is consistent with relativity.

Of particular importance here is the notion of an instant, which in ED involves several ingredients: (1) a spacelike surface $\sigma$ that codifies spatial relations and provides a criterion of simultaneity, (2) a specification of the statistical state $\rho $ and $\phi $ that is sufficient for the prediction of future states, and (3) an entropic step in which we are compelled to update the statistical state at one instant to generate the state at the next instant; it is this requirement that makes dynamics come alive.
\paragraph{Spacetime kinematics}
Having established the notion of an instant as relying on the concept of a surface $\sigma$, it is useful to provide an overview of some spacetime kinematics. We deal here with a background spacetime $\mathcal{M}$ with events labeled by the spacetime coordinates $X^{\mu}$ ($\mu=0,1,2,3$), and spacetime metric $g_{\mu\nu}(X^{\mu}$). We now foliate $\mathcal{M}$ into a sequence of spacelike surfaces $\{\sigma\}$, where each such three dimensional surface comes labeled with coordinates $x^{i}$ ($i=1,2,3$). The embedding of these surfaces is accomplished by the introduction of four embedding, or \textit{surface} variables $X^{\alpha }(x)\equiv X_{x}^{\alpha }$ that act as scalar functions on the surface $\sigma$. The role of these variables is crucial; they specify the location of the point $x$ on a particular surface $\sigma$.

When the spacetime geometry is itself fixed, the surface variables can be used to determine the metric induced on a surface due to the background metric $g_{\mu\nu}$. This is done in the standard way according to
\begin{equation}
g_{ab}=g_{\alpha \beta}X^{\alpha}_{a}\, X^{\beta}_{b},
\label{Induced metric}
\end{equation}
where $X^{\alpha}_{a} = \partial X^{\alpha}/\partial x^{a}$ projects spacetime tensors onto the surface $\sigma$.

If one wishes to pass from an initial surface $\sigma$ to the next
surface $\sigma^{\prime}$, then this can be accomplished by \textit{deforming} the surface $\sigma$ into $\sigma^{\prime}$ by means of the (infinitesimal) deformation $\delta \xi
_{x}^{A}=(\delta \xi _{x}^{\bot },\delta \xi _{x}^{a})$. The deformation $\delta\xi_{x}^{A}$ brings the point $x$ in $\sigma$ to the very same point $x$ in $\sigma$ by ``pushing" the surface $\sigma$, in spacetime, an amount $\delta\xi_{x}^{\perp}$ normal to $\sigma$ and an amount $\delta\xi_{x}^{a}$ parallel (tangential) to the surface. Note that, by convention, a deformation is defined by its components so that we can speak unambiguously about applying the same deformation $\delta\xi^{A}$ on different surfaces $\sigma$.
\paragraph{Duration}
In ED, time is defined to be simple, and useful. One place where viewpoint is implemented is with regards to the notion of separation between instants, or duration. That is, in ED we define time so that the dynamics appears as simple as possible. For short steps, the dynamics is dominated by the fluctuations. Thus a natural way to proceed is to define time as a measure of the fluctuations; as can be seen in (\ref{Fluctuations}), this can be accomplished by a prudent choice of the Lagrange multiplier $\alpha_{x}$.

The notion of duration that we seek here should be local. A natural choice is suggested by geometry. The geodesic drawn normal from the surface $\sigma$ to the subsequent surface $\sigma^{\prime}$ is the local proper time; which is the same as the normal component $\delta\xi^{\perp}_{x}$ of the deformation $\delta\xi_{x}^{A}$. This has all the properties we desire in a local notion of duration. This can be implemented by choosing the Lagrange multiplier to be
\begin{equation}
\alpha _{x}=\frac{1}{\hbar \delta\xi^{\perp}_{x}}\quad\text{so that}\quad
\left\langle \Delta w_{x}\Delta w_{x^{\prime }}\right\rangle =\frac{\hbar
\,\delta\xi^{\perp}_{x}}{g_{x}^{1/2}}\,\delta _{xx^{\prime }}.
\label{Duration}
\end{equation}%
With this, the transition probability $P\left[ \chi ^{\prime }|\chi\right]$ describes a type of Wiener process evolving in spacetime.
\paragraph*{Ordered instants}
Having established what is meant by an instant, we pass to the question of how these instants are ordered; this is, after all, the essence of dynamics. The main idea is this: an instant --- an informational state --- will be identified as \textit{posterior}, if it has been updated, relative to a \textit{prior} distribution that has not been. In ED the new information available to us is supplied in the form of the constraints (\ref{Constraint 1}) and (\ref{Constraint 2}), or equivalently, in the transition probability $P[\chi^{\prime}|\chi]$. To integrate new information into our state of knowledge we require a protocol for doing so.

To this end, the joint distribution $P[\chi^{\prime},\chi] = P[\chi^{\prime}|\chi]\,\rho[\chi]$ that expresses our uncertainty in the configuration $\chi$ and the subsequent configuration $\chi^{\prime}$. By application of the ``sum rule" of probability theory to the joint distribution we obtain
\begin{equation}
\rho ^{\prime }\left[ \chi ^{\prime } \right] =\int D\chi \, P\left[ \chi ^{\prime },\chi\right]=\int D\chi \,P\left[ \chi ^{\prime }|\chi\right] \,\rho \left[ \chi\right].
\label{Evolution equation}
\end{equation}%
This step requires no assumptions; it is true by virtue of the rules of probability theory. The assumption comes in the interpretation. As the distribution $\rho^{\prime}\left[ \chi ^{\prime } \right]$ is more ``up to date" than $\rho \left[ \chi \right]$, it is tempting to say that $\rho^{\prime}\left[ \chi ^{\prime } \right]$ is \textit{posterior} to $\rho[\chi]$, just as $\rho[\chi]$ is \textit{prior} to $\rho^{\prime}[\chi^{\prime}]$. Indeed, it is convenient to re-label $\rho \left[ \chi \right]$ as $\rho_{\sigma}\left[ \chi \right]$ and similarly label $\rho^{\prime}\left[ \chi ^{\prime } \right]$ as $\rho_{\sigma^{\prime}}\left[ \chi ^{\prime } \right]$. The posterior is obtained from the prior, and this defines an iterative process which generates a dynamics; the surfaces $\sigma$ acting as labels that monitor the progression.

This represents an important first step in achieving a covariant theory; we are no longer restricted to a single notion of an instant. Indeed, if we wish to express things differently, we can. We simply choose a different foliation. This amounts in the standard relativistic perspective to simply performing a local Lorentz transformation. (In the more familiar case of a flat background spacetime, i.e. Minkowski space, for example, the ability to choose a foliation is a proxy for performing a Lorentz boost.) And so, our approach here to entropic time has indeed provided us with the formal freedom of a relativistic theory. However this is not quite enough. One would, in addition, desire for the dynamics itself to be consistent with this relativistic freedom. For this we require an additional constraint; we must impose path independence. This is the subject of the subsequent sections.
\paragraph{Dissipative dynamics}
The integral updating protocol (\ref{Evolution equation}) and definition of duration (\ref{Duration}) together imply a set of local ``Fokker-Planck" (FP) equations, one per spatial point, given by
\footnote{The equations (\ref{FP equation}) are not true FP equations since they lack of a divergence in the configuration space $\mathcal{C}$ on the right-hand side of the equation. Comparing with the FP equation in \cite{Caticha Ipek 2014} we see that a spatial integral is absent in (\ref{FP equation}).}
\begin{equation}
\frac{\delta \rho_{\sigma}[\chi] }{\delta\xi^{\perp}_{x}}=-\,g_{x}^{-1/2}\frac{\delta }{\delta \chi _{x}}\left( \rho_{\sigma}[\chi]  \,\frac{\delta \Phi_{\sigma}[\chi] }{\delta \chi _{x}}\right),
\label{FP equation}
\end{equation}%
where we have introduced the functional $\Phi_{\sigma}[\chi]$ as
\begin{equation}
\Phi_{\sigma} \left[ \chi \right] =\hbar \,\phi_{\sigma} \left[ \chi \right] - \frac{\hbar }{2}\log \rho_{\sigma} \left[ \chi \right].
\end{equation}
These local FP equations describe the flow of probability from an initial surface $\sigma$ to a subsequent surface $\sigma^{\prime}$, where the current velocity $V_{x}\left[\chi\right]=\delta \Phi/\delta \chi_{x}$ determines the speed with which the probability flows. The functional $\Phi$ is of particular interest; it will later be identified with the Hamilton-Jacobi functional, or rather, the phase of the wave functional $\Psi$ in the quantum theory. 

To lay the ground for the forthcoming sections, we write the local time FP equation suggestively as
\begin{equation}
\frac{\delta \rho_{\sigma} \left[ \chi\right] }{\delta\xi^{\perp}_{x}}\equiv \frac{\tilde{\delta} \tilde{H}_{\perp x}\left[ \rho,\Phi \right] }{%
\tilde{\delta} \Phi_{\sigma} \left[ \chi \right] }.
\label{FP equation H}
\end{equation}
Here we have introduced $\tilde{\delta}/\tilde{\delta}\Phi_{\sigma}$, the \textit{ensemble} functional derivative of a quantity $\tilde{H}_{x}^{\perp }$.\footnote{Here we introduce the notion of an ensemble functional, or simply, e-functional. Just as a regular functional such as $\rho\left[\chi\right]$ maps a field $\chi$ into a real number (a probability in this case), an e-functional maps a functional, such as $\rho\left[\chi\right]$ or $\Phi\left[\chi\right]$, into a real number. Thus, just as one can define functional derivatives, one can also define ``e-functional" derivatives. Notationally, ensemble quantities recieve a tilde to distinguish themselves. See the appendix of \cite{Reginatto 2003} for a review of the ensemble calculus.} Note that eq.(\ref{FP equation H}) is not an assumption; a suitable $\tilde{H}_{\perp x}$ can always be found. In fact, one can check that
\begin{equation}
\tilde{H}_{\perp x}\left[ \rho ,\Phi \right] = \int D\chi\,g_{x}^{-1/2}\,\rho
\frac{1}{2}\left( \frac{\delta \Phi }{\delta \chi _{x}}\right) ^{2}+F_{x}\left[ \rho %
\right]   \label{e-Ham perp w/F}
\end{equation}%
where $F_{x}$ is an arbitrary ensemble functional of $\rho $, reproduces (\ref{FP equation}). In subsequent sections we will see that $\tilde{H}_{\perp x}$ can be cast in the role of a \textit{Hamiltonian} generator.
\section{Embeddability and path independence}
\label{Spacetime and consistency}
A covariant theory is composed of two similar but distinct aspects; one kinematical and the other dynamical. The former consists of the formal tools, i.e. the kinematics, of spacetime that serve the purpose of expressing Einstein's notion of (local) relative simultaneity. The dynamical aspect of the theory, on the other hand, is often not truly recognized. A great virtue of the canonical approach to covariant theories is that this rich dynamical structure is revealed. The work of DKT, and Kucha\v{r} and Teitelboim in particular, are important in this regard. Their work introduces a particularly elegant distinction between what constitutes the kinematics, and what principles underlie the dynamics of covariant theories.

Their approach was the familiar one. Slice spacetime into a foliation of spacelike surfaces. However, once one has foliated spacetime by a sequence of spacelike surfaces, the original four-dimensional diffeomorphism symmetry of spacetime appears to have been lost. It is only rediscovered by requiring foliation invariance, which is implemented by the purely \textit{kinematic} criterion of embeddability (see e.g. Ch.2 \cite{Teitelboim thesis} or \cite{Kuchar 1976 I}). The interpretation is rather clear: we must generate this foliation in a specific way such that the sequence of spatial slices mesh to form a spacetime. This condition of embeddability then manifests itself as an algebra of deformations.

This situation does not, however, speak at all to the dynamics. For this we require something more. The important contribution of Teitelboim and Kucha\v{r} was to recognize that this additional information was provided by the requirement of path independence: if the evolution from an initial to a final state can proceed through many paths, then all must agree. This is implemented by requiring that the generators of the dynamical theory form an algebra that closes in an identical fashion to that of the algebra of deformations. Our goal here is to first introduce the kinematic requirement of embeddability. We then provide an overview of the argument for path independence.
\paragraph{Embeddability}
Consider a functional $T[X^{\alpha}]$ of the surface variables $X^{\alpha}_{x}$ whose value is defined unambiguously by the surface that it resides on. We now consider a perturbation of that surface by a deformation $\delta\xi^{A} = (\delta\xi^{\perp},\delta\xi^{a})$,
\begin{equation}
\delta T[X] = \int dx \frac{\delta T[X]}{\delta \xi^{A}}\delta\xi^{A}.
\end{equation}
The derivative operators are given by
\begin{equation}
\frac{\delta}{\delta\xi^{\perp}} = n^{\alpha}_{x}\frac{\delta}{\delta X^{\alpha}_{x}}\quad\text{and}\quad \frac{\delta}{\delta\xi^{a}} = X^{\alpha}_{a}\frac{\delta}{\delta X^{\alpha}_{x}},
\end{equation}
where $n^{\alpha}_{x}$ ($n_{\alpha}n^{\alpha} = -1$) is the unit normal vector.

As shown in \cite{Kuchar 1976 I}, these operators form a set of non-holonomic basis vectors in hyperspace, the space of embedding variables. That is, while the operators $\delta/\delta X^{\alpha}$ all commute,
\[ \left[\frac{\delta}{\delta X^{\alpha}}, \frac{\delta}{\delta X^{\beta}}\right] = 0,\]
the operators $\delta/\delta\xi^{A}$ do not. Instead they are defined by non-vanishing commutators
\begin{equation}
 \left[\frac{\delta}{\delta \xi^{A}_{x}}, \frac{\delta}{\delta \xi^{B}_{x^{\prime}}}\right] = \int dx^{\prime\prime} \kappa^{C}_{AB}(x,x^{\prime};x^{\prime\prime})\frac{\delta}{\delta \xi^{C}_{x''}},
\end{equation}
where the $\kappa_{AB}^{C}$ play the role of structure ``constants."\footnote{These constants are not, in fact, constants. They depend on the metric of the surface, and thus, on the surface variables themselves through (\ref{Induced metric}).}

Embeddability is then a purely geometric requirement that the surfaces reside within the same spacetime. One can show (see e.g. \cite{Teitelboim 1972}) that this condition serves to fix the ``constants" $\kappa^{C}_{AB}$ uniquely. This is given explicitly as
\begin{subequations}
\begin{align}
[\frac{\delta}{\delta \xi_{x}^{\perp}},\frac{\delta}{\delta \xi_{x^{\prime}}^{\perp}}]&=-\left(g^{ab}_{x}\frac{\delta}{\delta \xi^{b}_{x}}+g^{ab}_{x^{\prime}}\frac{\delta}{\delta \xi^{b}_{x^{\prime}}}\right)\partial_{ax}\delta(x,x^{\prime})\label{Surface Deformations perp-perp}\\
[\frac{\delta}{\delta \xi^{a}_{x}},\frac{\delta}{\delta \xi_{x^{\prime}}^{\perp}}]&=-\frac{\delta}{\delta \xi_{x}^{\perp}}\,\partial_{ax}\delta(x,x^{\prime})\label{Surface Deformations tan-perp}\\
[\frac{\delta}{\delta \xi^{a}_{x}},\frac{\delta}{\delta \xi^{b}_{x^{\prime}}}]&=-\frac{\delta}{\delta \xi^{a}_{x^{\prime}}}\partial_{bx}\delta(x,x^{\prime})-\frac{\delta}{\delta \xi^{b}_{x}}\partial_{ax}\delta(x,x^{\prime}).\label{Surface Deformations tan-tan}
\end{align}
\end{subequations}
\paragraph{Path independence}
Now we turn to the main idea: path independence. The idea is that if there are two manners of updating, that they should agree. As a constraint on how one ought to update, path independence is a dynamical condition to be applied to any dynamical system that seeks relativistic covariance as a guiding assumption. The main assumption is that there exists some generator, say $G_{Ax}$, that generates an evolution for the dynamics.\footnote{Note that although we will focus main on Hamiltonian generators in this work, the generators $G_{Ax}$ need not be of this type. In fact, the requirement of path independence should hold for any generator that induces a dynamical flow that might be called covariant.}

The underlying logic is that we evolve the geometry and dynamical variables along two different paths and demand that these paths agree. That is, we can perform successive deformations $\delta \xi ^{A}$
then $\delta \eta ^{B}$, $\sigma \stackrel{\delta \xi}{\rightarrow}\sigma_{1}\stackrel{\delta \eta }{\rightarrow }\sigma ^{\prime }$ or execute them in the opposite order, $\sigma \stackrel{\delta \eta }{\rightarrow }\sigma_{2}\stackrel{\delta \xi }{\rightarrow }\sigma ^{\prime \prime }$. The key point is that the surfaces $\sigma ^{\prime }$ and $\sigma ^{\prime \prime }$
will generally differ and their difference is given by a third deformation $%
\delta \zeta ^{\alpha }$ which  takes $\sigma ^{\prime }$ to $\sigma
^{\prime \prime }$: $\sigma ^{\prime }\stackrel{\delta \zeta }{\rightarrow }%
\sigma ^{\prime \prime }$. Thus we can follow either of two paths which pass from $\sigma $ to $\sigma
^{\prime \prime }$: either we follow the direct path $\sigma \stackrel{\delta
\eta }{\rightarrow }\sigma _{2}\stackrel{\delta \xi }{\rightarrow }\sigma
^{\prime \prime }$ or we follow the indirect path $\sigma \stackrel{\delta
\xi }{\rightarrow }\sigma _{1}\stackrel{\delta \eta }{\rightarrow }\sigma
^{\prime }\stackrel{\delta \zeta }{\rightarrow }\sigma ^{\prime \prime }$. Consistency demands the two alternatives must agree.

The result is a set of conditions to be satisfied by the generators $G_{Ax}$, given by
\begin{equation}
[G_{Ax},G_{Bx^{\prime}}] = \int dx^{\prime\prime}\kappa^{C}_{AB}(x,x^{\prime};x^{\prime\prime})G_{Cx^{\prime\prime}},
\label{Path Ind G_Ax}
\end{equation}
where the bilinear product $\left[\cdot , \cdot\right]$ depends on the details of the generators. (In our current approach this bilinear product will be identified with a Poisson bracket.) The point, however, is not just that these generators form a sort of algebra. It is that this algebra, which pertains to the dynamical variables, must close in exactly the same manner as the generators of deformations, given in (\ref{Surface Deformations perp-perp})-(\ref{Surface Deformations tan-tan}). It is only those $G_{Ax}$ that satisfy these conditions that can be interpreted as generating a covariant dynamics.
\section{Non-dissipative covariant dynamics}
\label{Non-dissipative covariant dynamics}
The equations (\ref{FP equation}) collectively describe a diffusive dynamics for the distribution $\rho$. A curious diffusion in a background spacetime, but a diffusion nonetheless. We, however, desire a more vibrant dynamics that can also describe a quantum theory. To do this we must adjust and supplement the existing constraints so that not only $\rho$, but also $\Phi$ participates in the dynamics: as $\rho $ is updated so too must $\Phi $.

As such, we have here two distinct tasks. The first is to identify the appropriate information on the basis of which $\Phi$ will be updated. The other is to require the dynamics of both $\rho$ and $\Phi$ to satisfy the covariance requirement of path independence; this task, however, requires us to first choose a particular \textit{representation} of the generators $G_{Ax}$ that suits our purpose. We suggest here a solution that solves both problems simultaneously. Consider our solution in a step-by-step manner. First, drawing inspiration from the Hamiltonian ED in \cite{Bartolomeo et al 2014}\cite{Caticha Ipek 2014}, we propose that the joint dynamics of $\rho$ and $\Phi$ be formulated in the canonical framework such that their evolution is generated by a set of Hamiltonian generators $H_{A x}$. Moreover, we seek the simplest such theory. This implies that we join $\rho$ and $\Phi$ together as a canonical pair. 

This choice of dynamics has the added feature in that it singles out a preferred choice of representation for implementing path independence. That is, the generators $G_{Ax}$ appearing in (\ref{Path Ind G_Ax}) can be represented as Hamiltonian generators $H_{A x}$, with the brackets $[\,\cdot , \cdot \,]$ being identified with Poisson brackets. Naturally, a covariant dynamics for $\rho$ and $\Phi$ can then be achieved by choosing the Hamiltonian generators that satisfy the algebra (\ref{Path Ind G_Ax}). Implementing this is the subject of this section.
\paragraph{Hamiltonian formalism}
The key assumption is that the dynamical variables ($\rho,\Phi$) form a canonical pair. The implication is that the space of possible $\rho$'s and $\Phi$'s forms a phase space, called the ensemble phase space, or e-phase space for short. There is, however, also the issue of how to treat the surface variables that dictate the changes in the background geometry. The natural way to do this was proposed by Dirac \cite{Dirac 1951}.

Following Dirac, the surface variables will too be treated as if they were dynamical variables --- they are not. That is, we introduce by hand the auxiliary variables $\pi _{\alpha }(x)\equiv \pi _{\alpha x}$ that play the role of momenta conjugate to the $X^{\alpha}_{x}$'s. These $\pi $'s are defined strictly through the Poisson bracket (PB) relations, 
\begin{equation}
[ X_{x}^{\alpha },X_{x^{\prime }}^{\beta }]=0\,,\quad [ \pi_{\alpha x},\pi _{\beta x^{\prime }}]=0\,,\quad [ X_{x}^{\alpha },\pi
_{\beta x^{\prime }}]=\delta _{\beta }^{\alpha }\delta (x,x^{\prime }),
\label{PB HS}
\end{equation}
where the PBs are defined over the extended phase space with coordinates $(\rho,\Phi;X^{\alpha},\pi_{\alpha})$. This extended Poisson bracket is explicitly given as
\begin{equation}
\left[ G,F\right] =\int D\chi \,\left( \frac{\tilde{\delta} G}{\tilde{\delta} \rho }\frac{%
\tilde{\delta} F}{\tilde{\delta} \Phi }-\frac{\tilde{\delta} G}{\tilde{\delta} \Phi }\frac{\tilde{\delta} F}{\tilde{\delta}
\rho }\right) +\int dx\,\left( \frac{\delta G}{\delta X_{x}^{\alpha }}\frac{\delta F}{\delta \pi _{\alpha x}}-\frac{\delta G}{\delta \pi _{\alpha x}}%
\frac{\delta F}{\delta X_{x}^{\alpha}}\right) 
\end{equation}%
where $G$ and $F$ are arbitrary functionals of the geometry, $\rho$, and $\Phi $. The PB relations, thus, appear here in two strains, (1) to conveniently express the fixed background geometry in a consistent way, and (2), the ensemble-PB (e-PB) relations, which do indeed express genuine dynamics through the entropic updating of $\rho$ and $\Phi$.
\paragraph{Canonical Path Independence}
It is appropriate here to introduce the notion of a total Hamiltonian generator, which in a fixed background \cite{Dirac Lectures}\cite{Teitelboim thesis} has the form
\begin{equation}
H_{Ax} = \pi_{Ax} +\tilde{H}_{Ax}.
\end{equation}
Here we have introduced the quantities
\begin{equation}
\pi_{\perp x} = n^{\alpha}_{x}\pi_{\alpha x}\quad\text{and}\quad \pi_{a x} = X^{\alpha}_{ax}\pi_{\alpha x},
\end{equation}
which generate deformations for the surface variables, while it is $\tilde{H}_{Ax}$ that will be associated with the generators for $\rho$ and $\Phi$. 

The full tangential generator $H_{ax}$ can be decomposed completely
into separate ensemble and geometrical elements, $H_{ax}=\pi _{ax}+\tilde{H}_{ax}$, where $\tilde{H}_{ax}=\tilde{H}_{ax}$ is completely independent of the surface variables or $\pi$. The normal generator, on the other hand, does not cleanly separate in this way since $\tilde{H}_{\bot x}=\tilde{H}_{\bot x}\left[\rho,\Phi,X\right]$ contains the surface variables through the metric of the surface, $g_{ab}$.

Having identified a bilinear product, the PB, and a formal set of generators $H_{Ax}$ (yet to be explicitly determined), we can phrase the algebra (\ref{Path Ind G_Ax}) in the canonical language. A \textit{sufficient} (see e.g. \cite{Teitelboim 1972}\cite{Teitelboim thesis}) set of conditions for the generators $H_{Ax}$ to satisfy are given by
\begin{eqnarray}
\left[ H_{\bot x},H_{\bot x^{\prime }}\right]  &=&\left(
g_{x}^{ab}H_{bx}+g_{x^{\prime }}^{ab}H_{bx^{\prime }}\right) \partial
_{ax}\delta (x,x^{\prime })  \label{PB 1} \\
\left[ H_{ax},H_{\bot x^{\prime }}\right]  &=&H_{\bot x}\partial _{ax}\delta
x,x^{\prime })  \label{PB 2} \\
\left[ H_{ax},H_{bx^{\prime }}\right]  &=&H_{ax^{\prime }}\,\partial
_{vx}\delta (x,x^{\prime })+H_{bx}\,\partial _{ax}\delta (x,x^{\prime })
\label{PB 3}
\end{eqnarray}%
supplemented by the constraints 
\begin{equation}
H_{\bot x}\approx 0\quad \mbox{and}\quad H_{ax}\approx 0~.
\label{Hamil and Momen Constraints}
\end{equation}

The difference between this and the algebra (\ref{Path Ind G_Ax}) is the inclusion of these additional constraints (\ref{Hamil and Momen Constraints}). In a theory with a dynamical background geometry these so-called Hamiltonian constraints are fundamentally important as they contain entirely the dynamics of the theory. In a theory with a fixed background spacetime, however, these constraints arise purely due to the introduction of the additional superfluous variables $\pi_{\alpha x}$, which were brought in for convenience. Thus these constraints are not particularly important in our work here.
\paragraph*{Hamiltonian generators}
We now would like to discuss further the generators in our theory. In particular, we would like to identify the choices of the ensemble generators $\tilde{H}_{Ax}$. (The surface generators $\pi_{Ax}$ are already defined in terms of the canonical variables \cite{Dirac 1951}\cite{Dirac Lectures} of the theory, and so, require no extra discussion in this regard.) We begin with the tangential generators, as they are the simpler ones; they induce translations of the dynamical variables parallel to the surface. The change in $\rho $ or $\Phi $ under a tangential deformation $\delta \xi _{x}^{a}$
can be generated by the e-momentum,%
\begin{equation}
\tilde{H}_{ax}[\rho ,\Phi ]=\int D\chi \,\rho \lbrack \chi ]\frac{\delta
\Phi \lbrack \chi ]}{\delta \chi _{x}}\,\partial _{ax}\chi _{x}.
\label{e-Momentum}
\end{equation}%

We can compute the effect of $\tilde{H}_{ax}$ by using the PBs; the result should agree with the result obtained by Lie dragging the variables across the surface. Indeed, this can be done, and we obtain
\begin{equation}
\frac{\delta \rho[\chi] }{\delta \xi _{x}^{a}} =-\frac{\delta \rho[\chi] }{\delta \chi_{x}}(\partial _{ax}\chi _{x})=[\rho[\chi] ,\tilde{H}_{ax}]
\label{Functional Derivatives Tangential rho}
\end{equation}
and
\begin{equation}
\frac{\delta \Phi[\chi] }{\delta \xi _{x}^{a}} =-\frac{\delta \Phi[\chi] }{%
\delta \chi _{x}}(\partial _{ax}\chi _{x})=[\Phi[\chi] ,\tilde{H}_{ax}].
\label{Functional Derivatives Tangential Phi}
\end{equation}
One can then show that (\ref{e-Momentum}) does indeed satisfy the condition (\ref{PB 3}) with $\tilde{H}_{ax}$ in place of $H_{ax}$.

Of dynamical significance, and non-trivial to determine, is the normal generator $\tilde{H}_{\perp x}$, which generates the local evolution of $\rho$ and $\Phi$. One requirement that we ask of such a $\tilde{H}_{\perp x}$ is that applying it to $\rho$ should recover the local time FP equations (\ref{FP equation}); indeed, just such an e-functional was already introduced in (\ref{e-Ham perp w/F}) that accomplishes this. The problem of finding a covariant dynamics for $\rho$ and $\Phi$ then reduces simply to finding the appropriate $F_{x}[\rho]$ that satisfies (\ref{PB 1}).
\paragraph*{Solving the Poisson brackets}
The PB relations (\ref{PB 1})-(\ref{PB 3}) can be split up, almost completely, into pieces containing just the surface contributions $\pi_{Ax}$ and just the ensemble contributions $\tilde{H}_{Ax}$. One might say that the algebra decomposes into a product: surface $\otimes$ ensemble. In fact, for the PBs (\ref{PB 1}) and (\ref{PB 3}) this split is perfect,
\begin{align}
\left[ H_{\bot x},H_{\bot x^{\prime }}\right]  &= [\pi_{\perp x} ,\pi_{\perp x^{\prime}}] + \left[ \tilde{H}_{\bot x},\tilde{H}_{\bot x^{\prime }}\right]  \label{PB 1 split} \\
\left[ H_{a x},H_{b x^{\prime }}\right]  &= [\pi_{a x} ,\pi_{b x^{\prime}}] + \left[ \tilde{H}_{a x},\tilde{H}_{b x^{\prime }}\right]\label{PB 3 split}.
\end{align}
There is a slight plot twist, however, with (\ref{PB 2}). Since $\tilde{H}_{\bot x}=\tilde{H}_{\bot x}\left[ \rho ,\Phi ;X\right] $ also contains the surface variables $X^{A}_{x}$, we require both generators $\pi _{ax}$ and $\tilde{H}_{ax}$ to act on $\tilde{H}_{\bot x}$ in the appropriate fashion. The correct PB to be satisfied is thus
\begin{equation}
\left[ \pi _{ax}+\tilde{H}_{ax},\tilde{H}_{\bot x^{\prime }}\right] =\tilde{H}_{\bot
x}\partial _{ax}\delta(x,x^{\prime }).
\end{equation}
The issue of finding the correct generators then reduces to the problem of analyzing each set of generators separately.

When setting out to solve these equations for the generators, one notices that the decomposition of deformations into normal and tangential components has the pragmatic feature that it separates the dynamical contributions --- the normal deformations --- from the kinematic ones --- the tangential deformations. This is particularly apparent in solving the PBs (\ref{PB 2}) and (\ref{PB 3}), as these PBs refer merely to the geometrical properties of the variables involved.

For example, as we mentioned above, the tangential generator $\tilde{H}_{ax}$ can be computed directly through the use of Lie derivatives along the surface, which depends only on the surface geometry; no dynamics necessary. One can always find a suitable tangential generator that accomplishes this. (As it turns out, the tangential deformations along the surface form a true group and that the PBs (\ref{PB 3}) represents their algebra.) Thus the PBs (\ref{PB 3}) are uniquely satisfied by the $\tilde{H}_{ax}$  

The second set of PB relations, (\ref{PB 2}) is similarly easy to satisfy. The action of tangential deformations (\ref{e-Momentum}) is to just shift the the dynamical variables along the surface from a point $x^{a}$ to a nearby point $x^{a}+\delta \xi^{a}_{x} $. The PBs (\ref{PB 2}) then determine how a normal deformation behaves as it is dragged along the surface. Since we $H_{ax}$ is a covariant vector density, the PBs (\ref{PB 2}) determine merely that $H_{\perp x}$ is a scalar density; these conclusions are, of course, true for $H_{Ax}$, and thus for $\pi_{Ax}$ and $\tilde{H}_{Ax}$ individually.
\paragraph{Requiring consistency}
It is the relation (\ref{PB 1}) that poses extreme constraints on $\tilde{H}_{\perp x}$ and thus on the update of $\Phi$. The PBs (\ref{PB 1}) have the effect of restricting the form of $F_{x}$ introduced in (\ref{e-Ham perp w/F}).\footnote{Work along these lines has suggested that the degree to which (\ref{PB 1}) imposes constraints on $\tilde{H}_{\perp x}$ is indeed extreme, with the possibility that the only ``non-classical" potential being the quantum potential itself.} That is, the relation (\ref{PB 1}) is a condition on the allowed forms of $F_{x}[\rho]$.

In particular the PB relations \textit{require} that $\tilde{H}_{\perp x}$ have a specific form  
\begin{equation}
\tilde{H}_{\perp x}=\tilde{H}_{\perp x}^{0} + f_{x}\left[ \rho \right],
\label{e-Ham w/f}
\end{equation}%
where $\tilde{H}_{\perp x}^{0}$ is given by
\begin{equation}
\tilde{H}_{\perp x}^{0} = \int D\chi \rho \left\{ \frac{1}{2}\frac{1}{g^{1/2}_x} \left(\frac{\delta\Phi }{\delta \chi _{x}}\right) ^{2}+\frac{g^{1/2}_x}{2} g^{ij}\partial_{i}\chi _{x}\partial _{j}\chi _{x}\right\},
\end{equation}
and where $f_{x}$ is an arbitrary functional that must satisfy the condition
\begin{equation}
[\tilde{H}_{\perp x}^{0},f_{x^{\prime}}]=[\tilde{H}_{\perp x^{\prime}}^{0},f_{x}].
\label{Condition f_x}
\end{equation}
The dynamical evolution for the conjugate momentum $\Phi $ is then of the form%
\begin{equation}
-\frac{\delta \Phi }{\delta\xi^{\perp}_{x}}=\frac{1}{2}\frac{1}{g^{1/2}_x} \left(\frac{\delta
\Phi }{\delta \chi _{x}}\right) ^{2}+\frac{g^{1/2}_x}{2} g^{ij}\partial
_{i}\chi _{x}\partial _{j}\chi _{x}+\frac{\tilde{\delta} f_{x}}{\tilde{\delta} \rho }=[\Phi ,\tilde{H}_{\perp x}].
\end{equation}%
For a special choice $f_{x} = 0$ we recognize this equation to be a local time functional Hamilton-Jacobi (HJ) equation for the phase functional $\Phi$.
\section{Conclusion}
\label{Conclusion}
The theory is ``quantized" in a straightforward way by the appropriate choice of potential $f_{x}$, which is subject to the condition (\ref{Condition f_x}). One possible choice is 
\begin{equation}
f_{x}\left[ \rho \right] =\int D\chi \,\rho \,V_{x}\left(\chi_{x}\right)+\frac{\hbar ^{2}}{8\, g^{1/2}_x}\int D\chi \,\rho \left( \frac{%
\delta \log \rho }{\delta \chi _{x}}\right) ^{2},
\end{equation}%
where $V_{x}$ is a potential made up of self interactions of $\chi_{x}$. The other term is the so-called quantum potential. It is this latter contribution that leads to the patented linear evolution of quantum theory.

This connection to quantum theory is made obvious by combining the pair of dynamical variables $\rho[\chi]$ and $\Phi[\chi]$ into $%
\Psi[\chi] =\rho ^{1/2}e^{i\Phi /\hbar }$ and its complex conjugate $\Psi^{*}[\chi]$.\footnote{See \cite{Bartolomeo et al 2014} and \cite{Caticha Ipek 2014} for a more in depth discussion of this point.} By combining the local time FP and HJ equations we obtain the local time evolution for $\Psi$
\begin{equation}
i\hbar \frac{\delta \Psi }{\delta\xi^{\perp}_{x}}=-\frac{1}{g^{1/2}_x}\frac{\hbar ^{2}}{2}\frac{\delta ^{2}\Psi }{\delta \chi _{x}^{2}}+\frac{g^{1/2}_x}{2} g^{ij}\partial
_{i}\chi _{x}\partial _{j}\chi _{x}\,\Psi +V\Psi=i\hbar \left[ \Psi ,\tilde{H}_{\perp x }\right] 
\label{Local Schro Eqn}
\end{equation}%
which we recognize as a local time Schr\"{o}dinger equation. And, since the evolution of $\rho$ and $\Phi$ is constrained by the requirement of path independence, so too is that of $\Psi$, and so we have here a covariant quantum dynamics.
\paragraph*{Discussion}
The covariant quantum ED that we present here is an alternative to the standard Dirac method of quantization. The key feature is that we were able to obtain a theory that is quantum and statistical by implementing methods that were developed for classical theories. Thus our approach does not require any of the ad hoc rules of standard quantization and is free of the operator ordering ambiguities that plague the conventional quantization methods.

The chief benefit, however, of our approach is that of interpretation. In a covariant quantum theory, what is it that obeys the principle of relativity? Is it the ontic fields? Or is it the epistemic state $\Psi$? Our approach answers this in a definitive way in favor of the latter. It is our state of knowledge that is updated in a covariant manner, not the fields themselves. We expect this kind of clarity to become more important as we pursue an ED of gravity.
\subsubsection*{Acknowledgements}

We would like to thank D. Bartolomeo, C. Cafaro, N. Caticha, S. DiFranzo, A.
Giffin, D.T. Johnson, K. Knuth, S. Nawaz, M. Reginatto, C. Rodr\'{\i}guez,
and K. Vanslette for many discussions on entropy, inference and quantum
mechanics.

\end{document}